\documentclass[10pt, twocolumn]{article}
\usepackage{geometry}                
\usepackage{graphicx}
\usepackage{amssymb}
\usepackage{hyperref}

\title{Detection of the papermilling behavior}
\author{Igor Podlubny \\
		Technical University of Kosice, Slovakia
}
\date{November 20, 2025}                                          

\begin{document}
\maketitle

\begin{abstract}
{\small
Based on the analysis of the data obtainable from the Web of Science publication and citation database,
typical signs of possible papermilling behavior are described, quantified, and illustrated by examples. 
A MATLAB function is provided for the analysis of the outputs from the Web of Science.
A new quantitative indicator -- integrity index, or I-index -- is proposed for using it along 
with standard bibliographic and scientometric indicators. A case study is presented. }
\end{abstract}

\section{Introduction}

Two recent articles by Agricola et al. \cite{Agricola-etal-2025a,Agricola-etal-2025b} 
draw attention to the problems of scientific misconduct that during recent years 
affected the mathematical community. 

There is one important aspect, however, that they did not mention. All those 
fraudulent papers are submitted to journals by individuals pursuing not the mathematical discoveries, not the advancement of knowledge -- but obtaining financial profit, promotions, grants, positions in editorial boards, various awards, including being named ``highly cited researchers'' (HCR).

Based on the analysis of the data obtainable from the Web of Science publication and citation database, typical signs of possible papermilling behavior are described, quantified, and illustrated by examples. A MATLAB function is provided for the analysis of the outputs from the Web of Science. A new quantitative indicator -- integrity index, or I-index -- is proposed for using it along with standard bibliographic and scientometric indicators.

Identification of papers generated (or produced) by “paper mills” is an important problem. 
The description of the operation of “paper mills” can be found in \cite{Abalkina-2023} and at websites focusing on various forms of scientific misconduct, e.g.~\cite{Oransky-2024,Schneider-2020,Schneider-2022}.  

This paper, however, is devoted to the identification of the \emph{systematic papermilling behavior of individual persons}.  It appears that it is possible to tell conscientious scientists from the individuals whose only goal is the increase of their numbers of publications, citations count, and h-index. 

The paper is organized as follows. First, the problem with excluding Mathematics from the lists of highly cited researchers published yearly by Clarivate, is described, and it is explained how mathematics became an easy victim of papermilling. Then the visual signs of the papermilling behavior are described and quantified with the help of the MATLAB toolbox created for processing the publication and citation data downloaded from the Web of Science database. As a case study, the 2019 list of highly cited researchers in the field of mathematics is analyzed.

A new indicator, named the integrity index (I-index), is proposed with the aim to help differentiate conscientious scientists and those who exhibit the papermilling behavior. 

Although this study started with the close look at the field of mathematics as a victim of papermilling, the papermilling behavior can be observed in other fields, and examples of that are provided.  

Finally, the relationship between the introduction of article processing charges (APC) and the open access fees (OAF), on one side, and the observed massive papermilling behavior, on the other, is discussed.

\section{The good intentions...} 

The original idea of publications and citations indexes was to help scientists in their research work.
Eugene Garfield, the creator of the Science Citation Index, wrote~\cite{Garfield}:

\begin{quote}
\textit{``In this paper, I propose a bibliographic system for science literature that can eliminate the uncritical citation of fraudulent, incomplete, or obsolete data by making it possible for the conscientious scholar to be aware of criticisms of earlier papers.''}. 
\end{quote}

\noindent
However, there is a proverb that says that good intentions pave the road to hell, 
and the field of mathematics is currently an identified victim of massive scientific misconduct
related to manipulations with publications and citations counts considered 
as indicators of the quality of individual researchers. 
Clearly, other fields of science suffer similar problems.

\section{ \ldots and what happened}

In the description of the methodology for  preparing the Highly Cited Researchers\texttrademark\  2023 lists, 
Clarivate\texttrademark\ explains the reasons for excluding the field of mathematics from such list 
\cite[see section ``Exceptions and exclusions'']{HCR-methodology-2023}:

\begin{quote}
\textit{``We have chosen to exclude the Mathematics category from our analysis for this year.}

\textit{The field of Mathematics differs from other categories in ESI. It is a highly fractionated research domain, with few individuals working on a number of specialty topics. The average rate of publication and citation in Mathematics is relatively low, so small increases in publication and citation tend to distort the representation and analysis of the overall field. Because of this, the field of Mathematics is more vulnerable to strategies to optimize status and rewards through publication and citation manipulation, especially through targeted citation of very recently published papers which can more easily become highly cited (top 1\% by citation). This not only misrepresents influential papers and people; it also obscures the influential publications and researchers that would have qualified for recognition. The responsible approach now is to afford this category additional analysis and judgement to identify individuals with significant and broad influence in the field.''}
\end{quote}

It should be added that the lists of highly cited researchers in mathematics contained 
99 names in 2014, also 99 names  in 2015, 106 names in 2016, 
96 names in 2017, 90 names in 2018, 89 names in 2019, 70 names in 2020, 
74 names in 2021, 52 names in 2022, and no names after that (\cite{HCR-past-lists}).
These lists are based on the data from the preceding years -- 
for example, the 2022 list is based on the data from 2011 to 2021. 
In all the lists there are 243 unique names.
However, not all of them have unique Web of Science author identifiers. 

It is worth mentioning here, that the fact that publications and citations counts in mathematics are the lowest compared to other sciences is known, and has been used 
for comparing citation impact in different fields of science 
(\cite{Podlubny-2005,Podlubny-2011})
and for creating the first multidisciplinary list of highly cited researchers based on their citation counts normalized with respect to mathematics
(\cite{Podlubny-2006}).

At the time of writing of the paper  \cite{Podlubny-2006}, 
the Highly Cited  Researchers  list (at the time known as Thomson ISI) 
was compiled based on the citations counts over the duration of the entire career. 
There was also a category called ``Research front'' including 
those papers that received more than some threshold number of citations within short time 
(a couple of years only); indeed, this could indicate that such papers 
might be bringing really new or significant results.

Currently, Clarivate compiles its Highly Cited Researchers (HCR) lists 
using something similar to those previous ``Research front'' papers~\cite[see section ``Preliminary candidate selection'']{HCR-methodology-2023}:

\begin{quote}
\textit{
``For the Highly Cited Researchers 2023 analysis, 
the papers surveyed were those published during 2012 to 2022 
and which then ranked in the top 1\% by citations 
for their ESI\footnote{Essential Science Indicators\texttrademark\  -- another Clarivate product.} field and year (the definition of a highly cited paper).''}
\end{quote}

This means that instead of the accumulated whole-career citation impact, the ``beans''  
called ``highly cited papers'' are counted for a relatively short time window of ten years. 
Due to this change of  rules, it became possible -- exactly as Clarivate described  -- 
to arrange relatively  small numbers of citations for a set of several papers and, 
as a result, get a particular person to the HCR list. 
The field of mathematics appeared particularly 
vulnerable to this type of action, because of its low citations counts per paper. 
Indeed, mathematics has the intrinsic property of a usually long delay between 
the publication of a  result and its wide recognition and use, reflected 
in the number of citations. 

To artificially increase the number of citations for 
the produced/selected ``beans'' within a short time frame, in order to obtain various benefits, 
the manipulation of citations is  used, which is done by producing more insignificant papers with the sole purpose of increasing the citation counts for the chosen ``beans''. 
This can be done easily because of the Clarivate's imperfect methodology and the way of counting citations. 

According to Clarivate's in table of 1\% citation thresholds for mathematics, 
a paper published in 2025 is considered highly cited if it received 3 (three) citations, 
a paper published in 2024 needs 9 citations, 2023 paper -- 19 citations, 
2022 paper  -- 30 citations, 2021 paper  -- 42 citations, 2020 paper -- 56 citations, 
2019 paper -- 64 citations, 2018 paper -- 74 citations, 2017 paper-- 77 citations, 2016 -- 81 citations, 
and a 2015 paper needs 92 citations for being considered as highly cited. 
When counting these citations, Clarivate does not ignore self-citations by all co-authors,
and even if only the co-authors cited their joint paper in three other papers indexed by Clarivate, 
that joint paper becomes ``highly cited''. 
This methodology and low thresholds allows coordinated action of groups of researchers
for producing bunches of papers in co-authorship citing each other. 

Apparently, those people who join a papermilling group in order to increase their publication counts,  
also  become  the ``suppliers'' of citations for other members of the group  
through their subsequent papermilling publications. 
This leads to the \emph{snowball effect}, and is likely the reason for the rapid monotonic growth of publication counts of individuals involved in the papermilling process.  
It also  leads to the growth of the number of co-authors of such individuals.

\section{The data source}

The data for further processing are obtained by exporting search results 
from the Web of Science (WoS). The necessary steps are the following. 

\begin{itemize}
\item
Go to the Web of Science and open Search. 

\item
Switch  to the RESEARCHERS tab, and start typing the last name; in the appearing dropdown list select the necessary last name. Then, similarly, enter the first name (start typing and select from the dropdown list).  Eventually, it is possible to add name variants. After entering the surname and the name, click on the ``Search'' button. 

\item
A researcher may have one or more affiliations. 

\begin{itemize}
\item[(a)]
If a researcher has one affiliation, click on the name of the researcher.
\item[(b)]
If a researcher has/had more than one affiliation, 
select proper affiliations using the checkboxes, 
and click on the ``View as combined record'' button. 
\end{itemize}

\item
In the right pane of the next screen locate and click the ``View citation report'' button. 

\item
On the next screen  called ``Citation Report'', which provides an overview of the researcher's 
indicators (publications, citations, H-index, and some other details), click on the link ``Export Full Report'', and save all outputs (use the ``Records from:'' option) using the ``Excel File'' format. 

\end{itemize}

The above procedure allows exporting up to 1000 records, 
which sounds like a reasonable number. 
However, some researchers have more than 1000 publications indexed in the WoS, 
and in such cases it is necessary to use the ``Records from:'' option several times with subsequent manual merging of the data from the obtained spreadsheet files.

\section{\raggedright Papermilling behavior:  visual signs} \label{sec:example}

On the Web of Science ``Citation Report'' page there is also the plot titled 
``Times Cited and Publications Over Time''. 
The vertical axis on the left is for publications counts, and
the vertical axis on the right is for citations counts; these two scales are different. 

\begin{figure}[t]
\center
\includegraphics[width=\columnwidth]{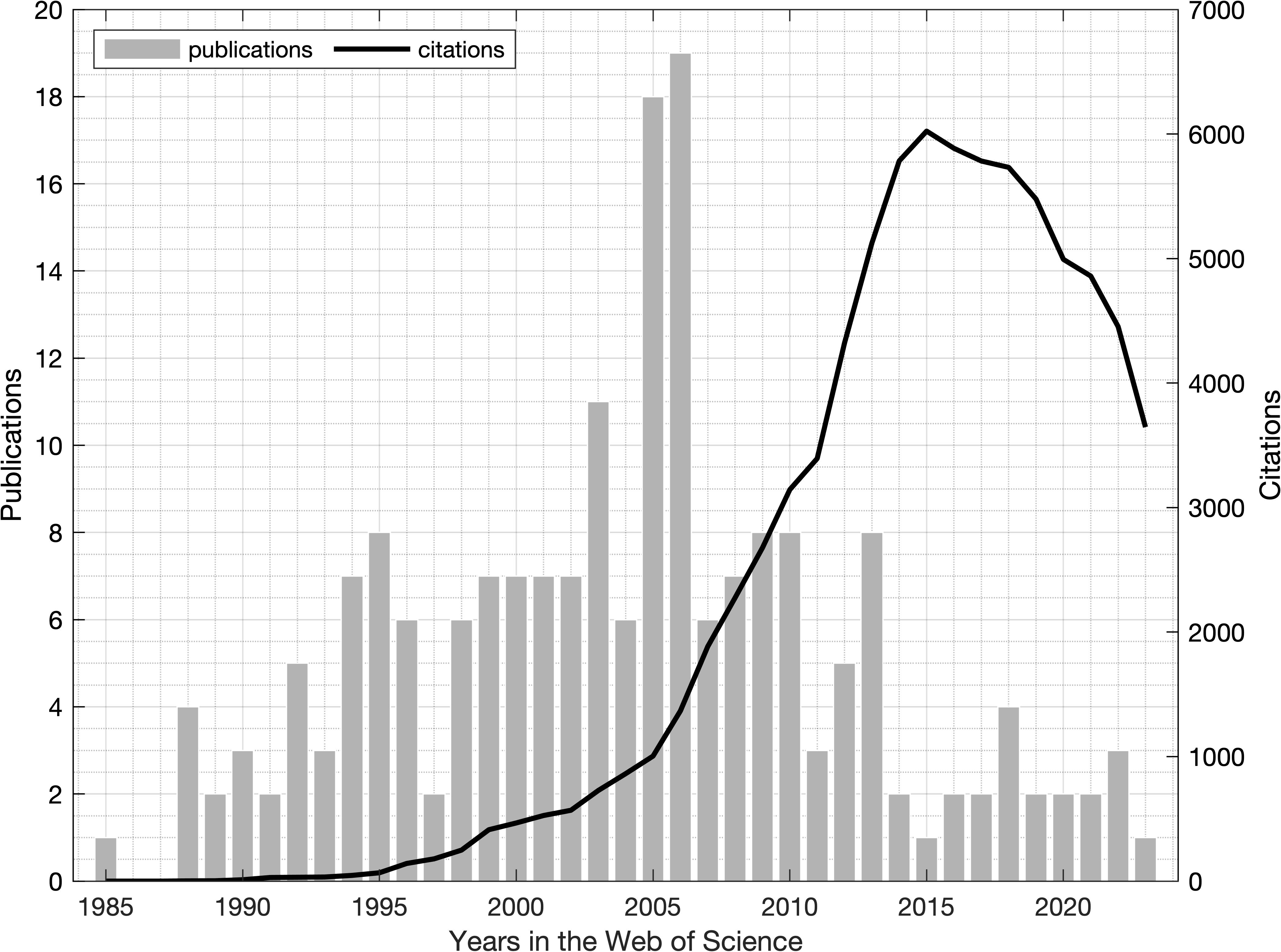}
\caption{Typical performance of a conscientious mathematician.}\label{fig:normal-example}
\end{figure}

\begin{figure}[t]
\center
\includegraphics[width=\columnwidth]{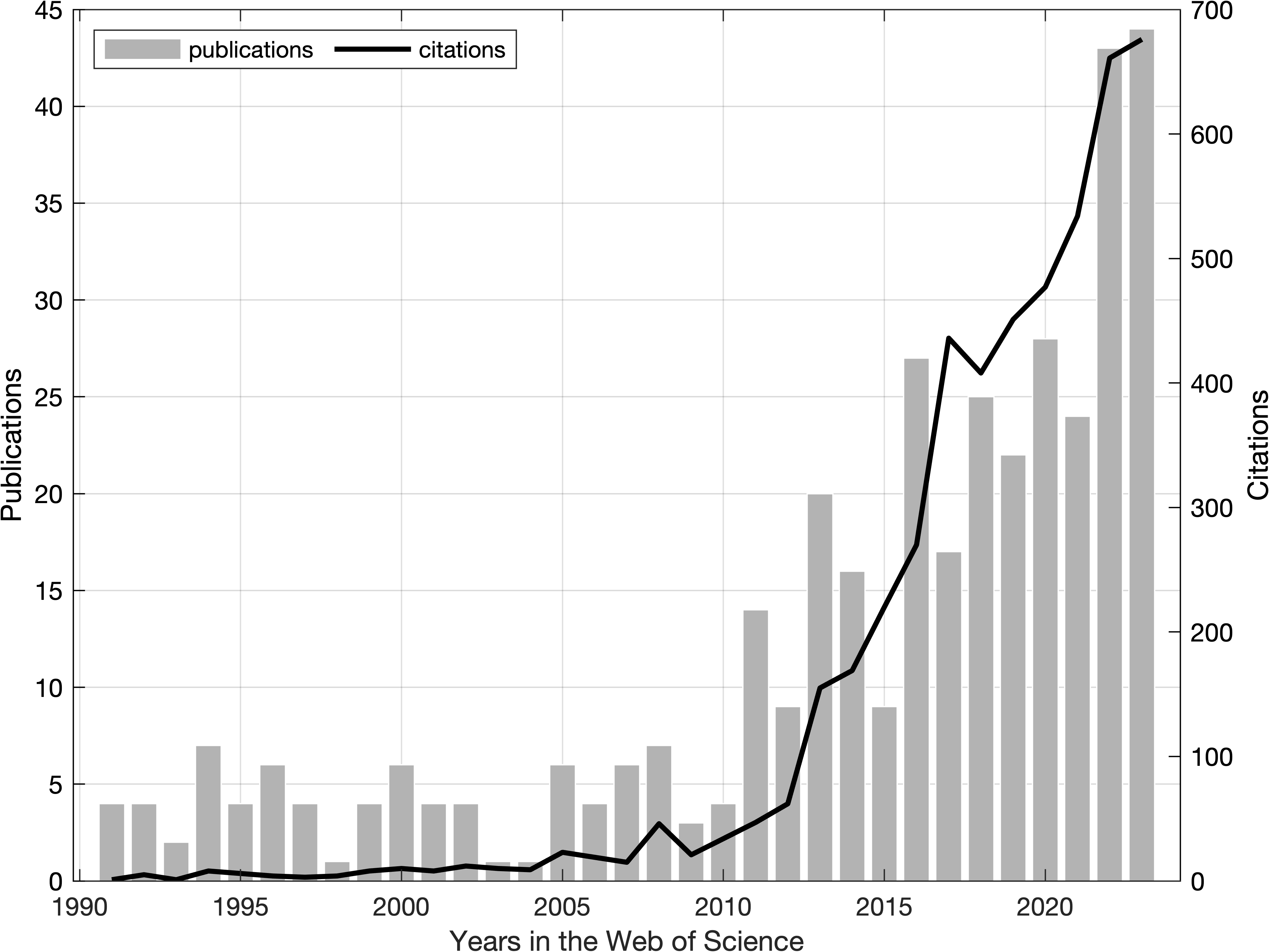}
\caption{Typical visual signs of possible papermilling in mathematics.}\label{fig:papermiller-example}
\end{figure}

This plot allows for hovering over or clicking on the vertical bars (or on the line) 
in order to see the number of publications and citations received in a particular year.

Two sample outputs, produced by the MATLAB toolbox described below in this article,
are shown in Fig.~\ref{fig:normal-example} (researcher R1) and Fig.~\ref{fig:papermiller-example} (researcher R2),
and in both cases the researchers are mathematicians 
who have been listed in the Highly Cited Researchers lists in the past; 
their career lengths are similar.

Researcher R1 had 196 papers indexed in the Web of Science, 
with at most 19 items in one year.
The publication activity of R1 was slightly increasing before 2011, 
reaching 18 and 19 publications in 2005 and 2006, after which it started declining. 
The number of citations of R1 had been growing with a significant delay, 
reaching its peaks in 2015.
The number of R1's papers  before 2011 was 159, with 128 co-authors from 15 countries.
In the period of 2011-2023, the number of publications of R1 was 37, with 36 co-authors from 6 countries.

Researcher R2 had 384 papers indexed in the Web of Science, 
with as many as 45 paper in one year. 
The publication activity of R2 before 2011 was low, with about 4 papers in one year, 81 papers in total, 
and with little citation impact; 
the number of co-authors of R2 before 2011 was 17, from 12 countries.
   
However,  from 2011 on, the number of publications of R2 started to grow rapidly, 
giving the total of 303 publications in 2011--2023, reaching the peak of 45 papers in 2022, 
and the number of citations started to grow synchronously with the number of publications. 
The total number of co-authors of R2 during 2011--2023 was 148, from 35 countries.

The described differences between researchers R1 and R2 are obvious. 
It should be mentioned that the publication pattern of R1 also reflects 
the generally known  observation that significant results are achieved 
in the first half of a mathematician's career, and that the publication activity 
normally decreases with age. 
In the case of the researcher R2, one can observe the aforementioned \emph{snowball effect}
typical for the papermilling behavior.

Obviously, in the case of researcher R2, authoring 45 papers in mathematics in one year is not typical,
and the synchronous growth of citation counts alongside the growth of paper counts is also atypical. 
On the contrary, in the case of researcher R1, the continuous growth of citations counts
despite the decrease in the number of papers is normal and shows the long-lasting
impact of R1's works published in the first half of the considered time period.

The following features in patterns of publications and citations counts are possible indicators of the papermilling behavior:
\begin{itemize}

\item
for a significantly long time period of career, the number of publications 
and the number of citations of a researcher are low; 

\item 
starting from some particular year, the number of publications starts to grow rapidly  
(usually monotonically), and reaches such numbers of papers per year that are unrealistic for a conscientious and responsible scientist; 

\item
the number of citations exhibits the same rapid growth behavior starting practically at the same time; 

\item
there is no delay between the rapidly growing number of publications 
and the rapidly growing number of citations -- they increase simultaneously;

\item
the number of co-authors (and their affiliations and countries) 
notably increases when the number of publications starts increasing. 

\end{itemize}

Besides these symptomatic signs of the overall publishing behavior, 
there are also symptoms that require scrutiny of each particular publication.

    	First, papermilling articles do not bring new ideas, 
their contribution is minimal (if any),
and they are often simply incorrect.

	Second, they usually include longer lists of references (with many unnecessary references) 
and often use bulk citations in the text, 
like \emph{``see [17--22, 34--38]''} ,  or \emph{``this was studied in [2]--[15]''}. 
Alternatively, many individual sentences in the introduction contain just one or two references,
so the bulk citations are spread across some usually unnecessary text.

\section{\raggedright Basic quantitative indicators of the papermilling behavior}

The number of publications in each year and the number of citations in the same year
are two time series, and we can investigate their properties and mutual relationship. 
To quantify the previous observations, the following indicators can be used when examining 
the publication activity of researchers:
\begin{itemize}
\item
correlation between the publications counts and the citations counts; 
high value of the correlation coefficient indicates possible papermilling;

\item
the delay between the citations counts and the publications counts; 
if the correlation is high and the delay is small, then 
this indicates possible papermilling; 

\item
unusually large numbers of publications in one year, with increasing trend,
indicate possible papermilling;

\item
low integrity index (see the next Section). 

\end{itemize}

\section{Integrity index}\label{i-index}

The main goal of a conscientious scientist is to discover new knowledge, 
and not to increase the number of his/her publications.
In this respect, it is worth recalling Theodore von K{\'a}rm{\'a}n's 
approach to comparing scientists~\cite[page 4]{Karman}:
\begin{quote}
\textit{
“If you define a great scientist as a man with great ideas, 
then you will have to rate Einstein first -- he had four great ideas. 
In the history of science perhaps only Sir Isaac Newton is ahead of Einstein, 
because he had five or six ideas. 
All the other major scientists of our age are associated with just one, 
or at the most two great ideas.''}
\end{quote}

The growing numbers of published papers make it increasingly difficult to distinguish 
the valuable, novel, and original material from the materials that just increase
 the global number of publications. 

The motivation for introducing the h-index by Jorge Hirsch~\cite{Hirsch-PNAS-2005}
was to focus on the publications 
which received sufficient attention of the related scientific community reflected in citations, 
and not to focus on large numbers
of publications that did not draw notable attention.
However, the papermills allow a fast increase of the h-index 
by  generating a large number of worthless papers and citations to such papers,
also in worthless articles; this is the aforementioned \emph{snowball effect}.

The solution can be  introducing a quantitative indicator, which would implement 
von K{\'a}rm{\'a}n's idea by lowering the motivation for increasing the number of publications
by penalizing the unnecessary increase of the total number of publications.

As such indicator, I suggest the ratio of the h-index and the total number of publications. 
This can be considered as a kind of a signal-to-noise ratio, 
called \emph{the integrity index}, and denoted as ``$I$-index''. 

If one author with h-index $H=32$ has 64 papers indexed in the Web of Science 
and the other author with the same h-index $H=32$ has 320 papers, 
then the integrity index of the first author is $I_1 = 32/64 = 0.5$
and the integrity index of the second author is  $I_1 = 32/320 = 0.1$. 
This means the higher probability that the first author focuses on
publishing novel original results and not on just publishing more papers.

In the example discussed above, 
the integrity index of researcher R1 is $I_1 = 86/196 = 0.44$, 
while the integrity index of researcher R2 is $I_2 = 34/384 = 0.08$. 
The integrity index can be used to differentiate conscientious scientists 
from the possible papermillers. 
In the case of the papermilling behavior, 
the integrity index is very low compared to scientists 
with the similar values of h-index.

\section{The MATLAB function}

A function for MATLAB, ``researcherprofilewos.m'',  has been created (\cite{MATLAB4WOS}) to simplify analysis of the data obtained from the Web of Science.  
All figures in this article are produced with the help of this tool.

This function takes an Excel file obtained from the Web of Science, 
and produces the following basic outputs:
\begin{itemize}
    \item 
    coefficient of linear correlation between publications and citations;
    \item 
    delay between citations and publications (only meaningful if the correlation is strong);
    \item 
    maximal number of papers in one year;
    \item 
    h-index (taken from the WoS);
    \item 
    $I$-index (integrity index);
    \item 
    total number of publications;
    \item 
    total citations count;
\end{itemize}
and the following additional information:
\begin{itemize}
    \item 
    starting year for publications;
    \item 
    minimal number of papers in one year;
    \item 
    average number of papers in one year;
    \item 
    average number of citations per paper.
\end{itemize}

The most important output is the coefficient of linear correlation between publications and citations.
Strong correlation means that the citation dynamics is strongly linked 
to the publication dynamics. 
In the case of researcher R2 the coefficient of linear correlation 
between publications and citations is $r_2=0.94$, and there is no delay between 
the growth of the number of publications and the number of citations. 
For researcher R1, the correlation coefficient is the negative number $r_1=-0.23$, 
which means that  R1's citation counts grow despite decreasing publications counts.

\section{A case study}

As a case study, let us take a look at the list of 89 researchers identified by Clarivate as highly cited researchers in the field of mathematics and considered by Agricola et al. \cite{Agricola-etal-2025a}. 
 
\begin{figure}[t!]
\center
\includegraphics[width=\columnwidth]{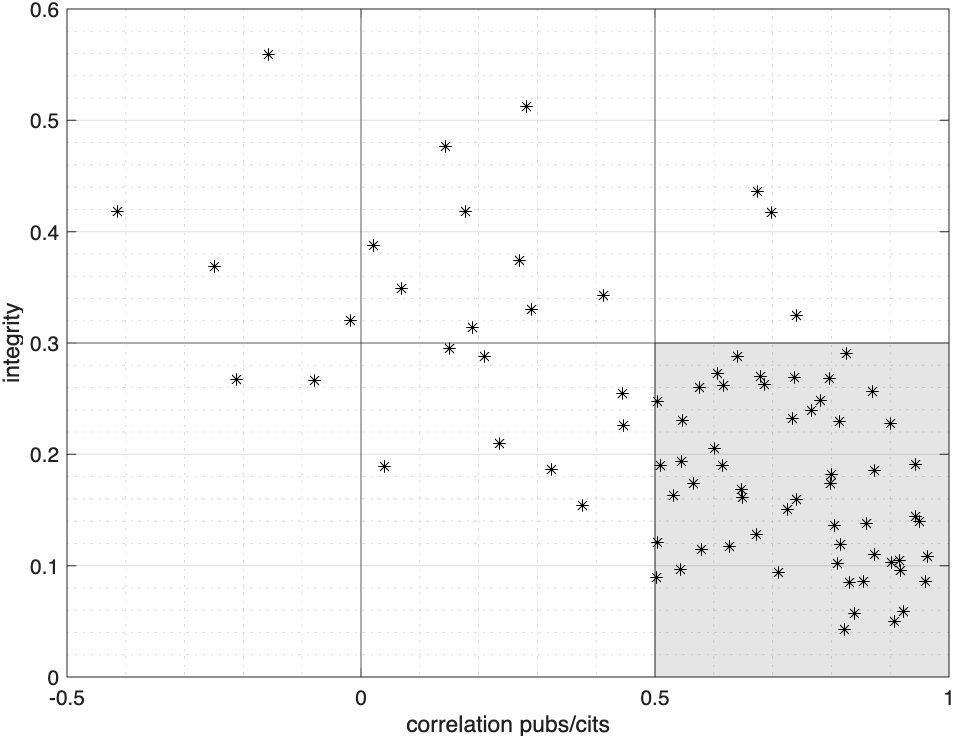}
\caption{Integrity index versus correlation between the time series of papers and citations}\label{fig:corr-integrity}
\end{figure}

\begin{figure}[t!]
\center
\includegraphics[width=\columnwidth]{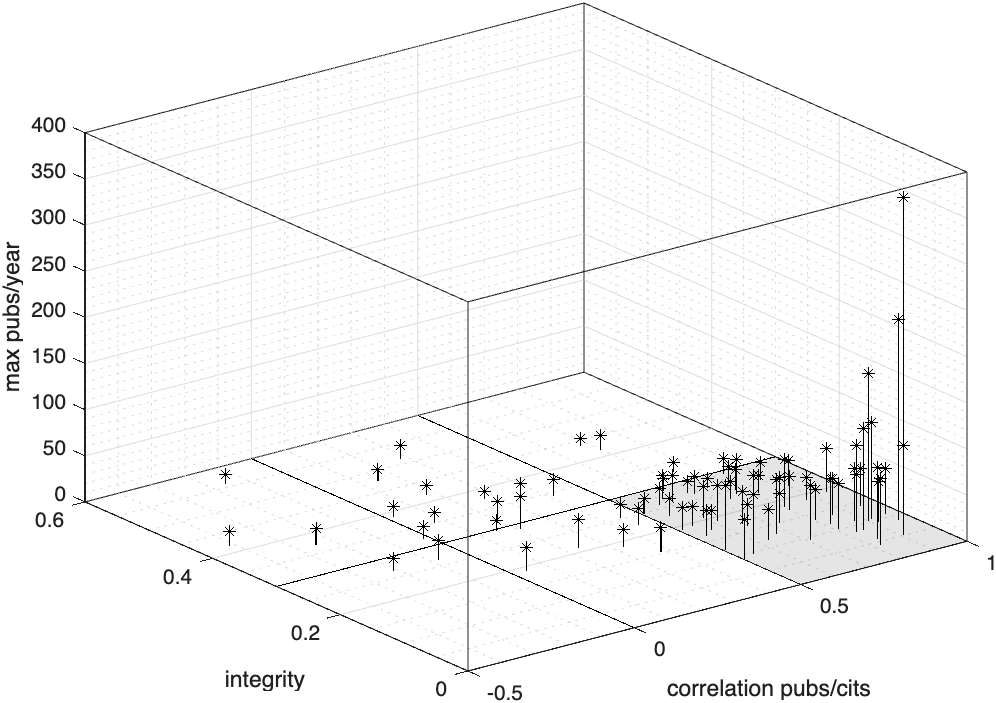}
\caption{Integrity, correlation between papers and citations, 
and maximum number of papers in one year.}
\label{fig:corr-integrity-maxp}
\end{figure}

\begin{figure}[t!]
\center
\includegraphics[width=\columnwidth]{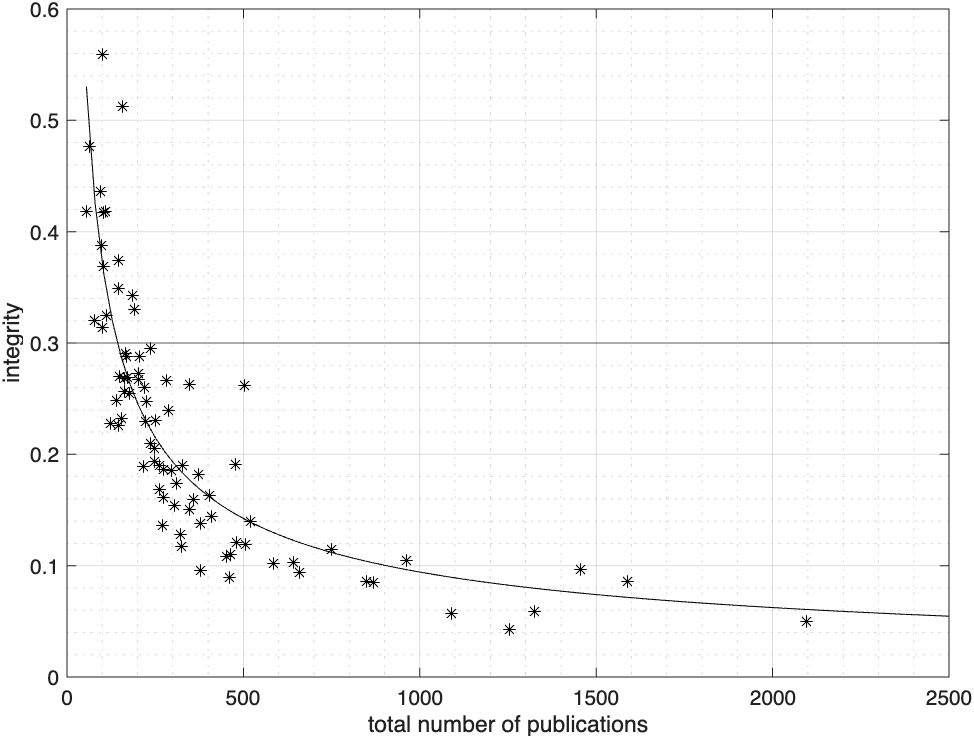}
\caption{Integrity index versus total number of papers, along with the power-law fit.}
\label{fig:integrity-pubs}
\end{figure}

\begin{figure}[t!]
\center
\includegraphics[width=\columnwidth]{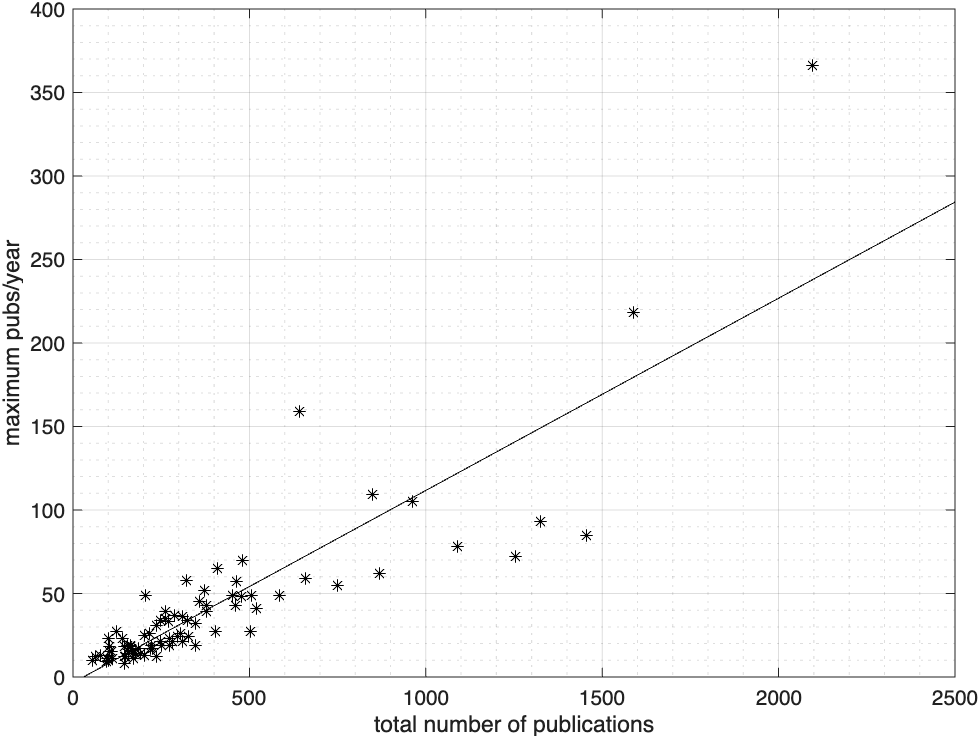}
\caption{Maximum number of papers in one year versus total number of papers, along with the linear fit.}
\label{fig:maxp-pubs}
\end{figure}

Unexpectedly, not all of them could be uniquely identified in the Web of Science by their Web of Science author identifiers\footnote{This is strange, since inclusion of a researcher in the HCR list is considered as a distinction and must be always verifiable. This is especially important when there are numerous profiles with the same or very similar full names. \par Also, some of the researchers, who are listed in HCR 2019 in the field of mathematics, do not have such a mark in their profiles in the Web of Science.}. 
The data for the remaining 82 researchers were processed using the aforementioned MATLAB function, and the  results are presented in Figs.~\ref{fig:corr-integrity}--\ref{fig:maxp-pubs}.

One of the main reasons for the papermilling behavior is the increase of the h-index, 
and the increase of the number of papers and the number of citations for this purpose results in strong correlation between the time series of publications and the time series of citations, practically without noticeable delay. As a result of inflating the total number of publications, the integrity index decreases. This is shown in Fig.~\ref{fig:corr-integrity}.

The shaded rectangle corresponds to correlation greater than 0.5 and integrity index less than 0.3, and  68\% of the 82 data points belong to this region. For the points in this region, the average total number of publications is around 487, the average maximum number of papers published in one year is around 52, and the average number of paper in one year is  around 16.  

For the data points outside the gray rectangle, 
the average total number of publications is around 159 (three times less), 
the average maximum number of papers published in one year is around 17 (also three times less), 
and the average number of paper in one year is  around 6 (2.7 times less).

There are only three data points with high publications/citations correlation and integrity index greater than 0.3, but the corresponding total paper counts are around 100, and the maximum number of papers in one year is around 12.

In Fig.~\ref{fig:corr-integrity-maxp}, the the maximum numbers of papers in one year are shown for the points from Fig.~\ref{fig:corr-integrity}. This shows that the large numbers of papers in one year are typical for a large group of researchers with high correlation between their publications and citations time series.

From the Fig.~\ref{fig:integrity-pubs}, it is obvious that increasing the total number of publications means the decrease of the integrity index, and this relationship is very well fit by the power law.
For the considered group of researchers, the equation of the fitting curve is 
\hbox{$I(p) = 1.7524 \, p^{-0.5956}$}, where $I(p)$ is the integrity index, and $p$ is the total number of papers.

The linear fit of the relationship between the total number of papers and the maximum number of papers in one year, shown in Fig.~\ref{fig:maxp-pubs}, confirms that the papermilling behavior leads to overproduction of publication units.  
For the considered group of researchers, the equation of the fitting line is
$m(p) = 0.1151\, p  -3.3981$, where $m(p)$ is the maximum number of papers in one year, 
and $p$ is the total number of papers.

\section{\raggedright Reasons for the papermilling behavior}

There are various reasons for papermilling, and being listed in the lists of highly-cited researchers (HCR)
is only one of them. The other reason is financial, because many universities link salaries and 
other financial benefits to the numbers of publications and citations, rankings, and also HCR listings. 
One of the most widespread reasons is  pursuing  promotion. 
Usually a candidate aspiring for promotion has to have a certain number of publications 
and citations in journals and conferences indexed in Web of Science or Scopus databases. 
The desire to achieve a promotion in a short time encourages some researchers to engage in papermilling activity and collaboration,
 because doing conscientious research and waiting for the eventual 
future citation impact delays the desired promotion. 

\begin{figure}[p]
\center
\includegraphics[width=\columnwidth,height=35ex]{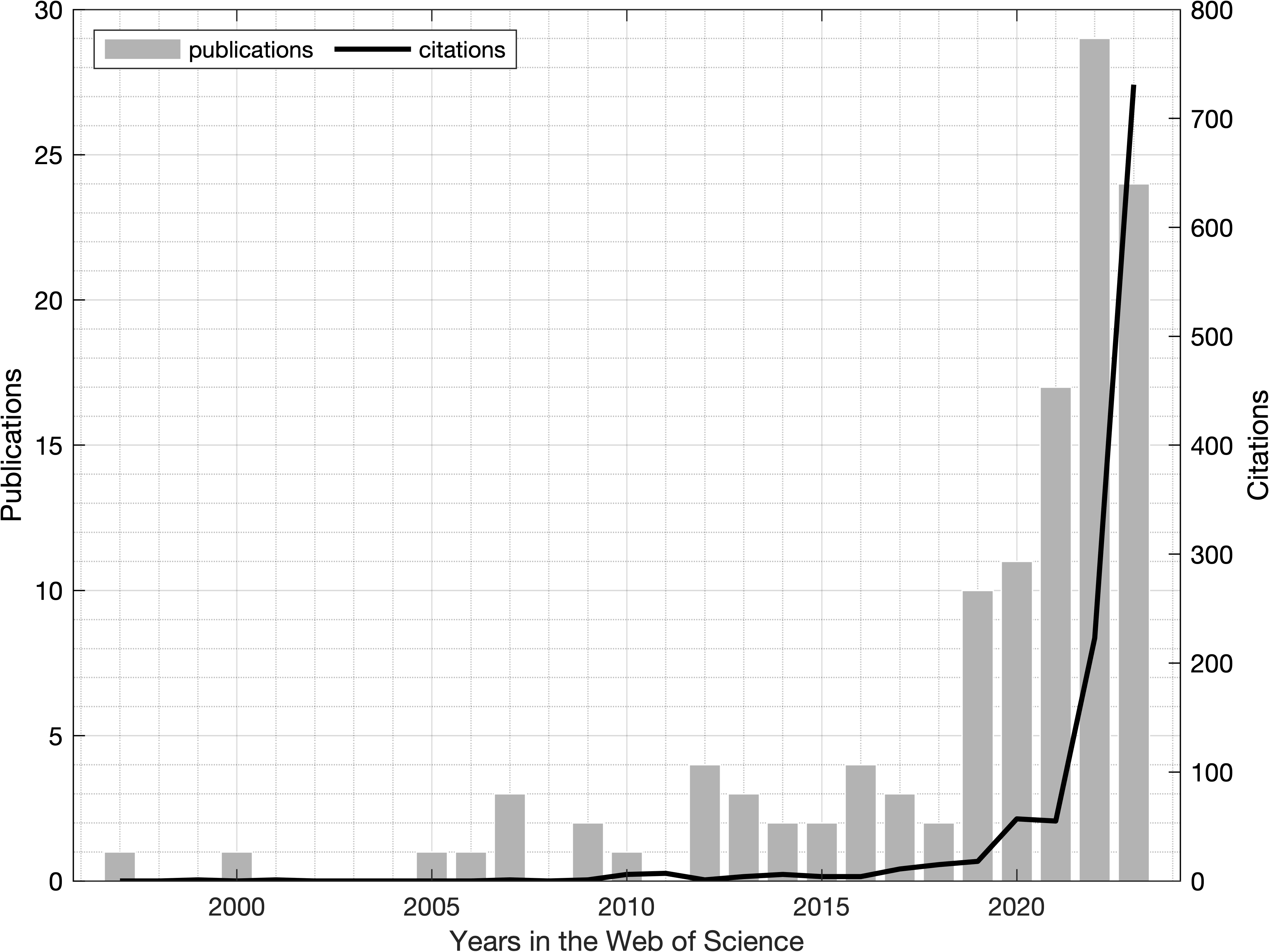}
\caption{Papermilling for promotion: \\ example~1 (mathematics, CZ).}\label{fig:promotion1}
\end{figure}

\begin{figure}[p]
\center
\includegraphics[width=\columnwidth,height=35ex]{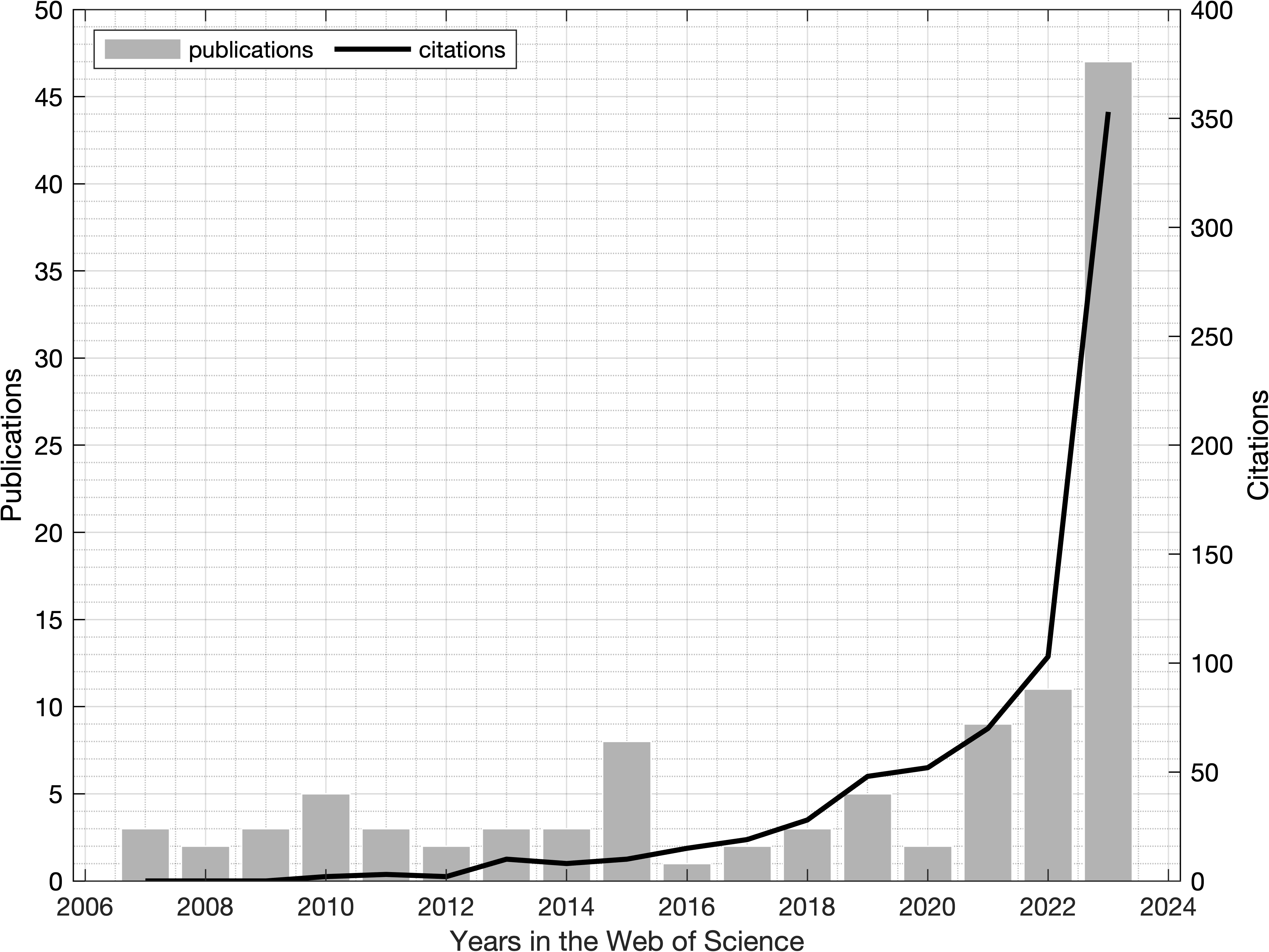}
\caption{Papermilling for promotion: \\ example~2 (environmental science, SK).}\label{fig:promotion2}
\end{figure}

Two probable examples of such kind 
are shown in Fig.~\ref{fig:promotion1} (mathematics, Czech Republic; 
correlation between publications and citations is $r=0.9318$, maximum number of papers in one year is 28, total number of publications is 120, integrity index $I=0.15$)
and Fig.~\ref{fig:promotion2} (mixed bag of environmental science, business, etc., Slovakia;
correlation between publications and citations is $r=0.99691$, maximum number of papers in one year is 48, total number of publications is 142, integrity index $I=0.1479$).

The damage done by the promotion-motivated papermilling should not be underestimated.
This type of behavior is contagious and may strongly affect colleagues and students,
pushing them to similar wrong and unethical practices. 
In addition, the promotion-motivated papermilling frequently helps such papermillers to get to influential positions or committees where they make important decisions about hiring, grants, and promotions of honest and conscientious researchers 
 -- and also of researchers similar to themselves.

\begin{figure}[p]
\center
\includegraphics[width=\columnwidth]{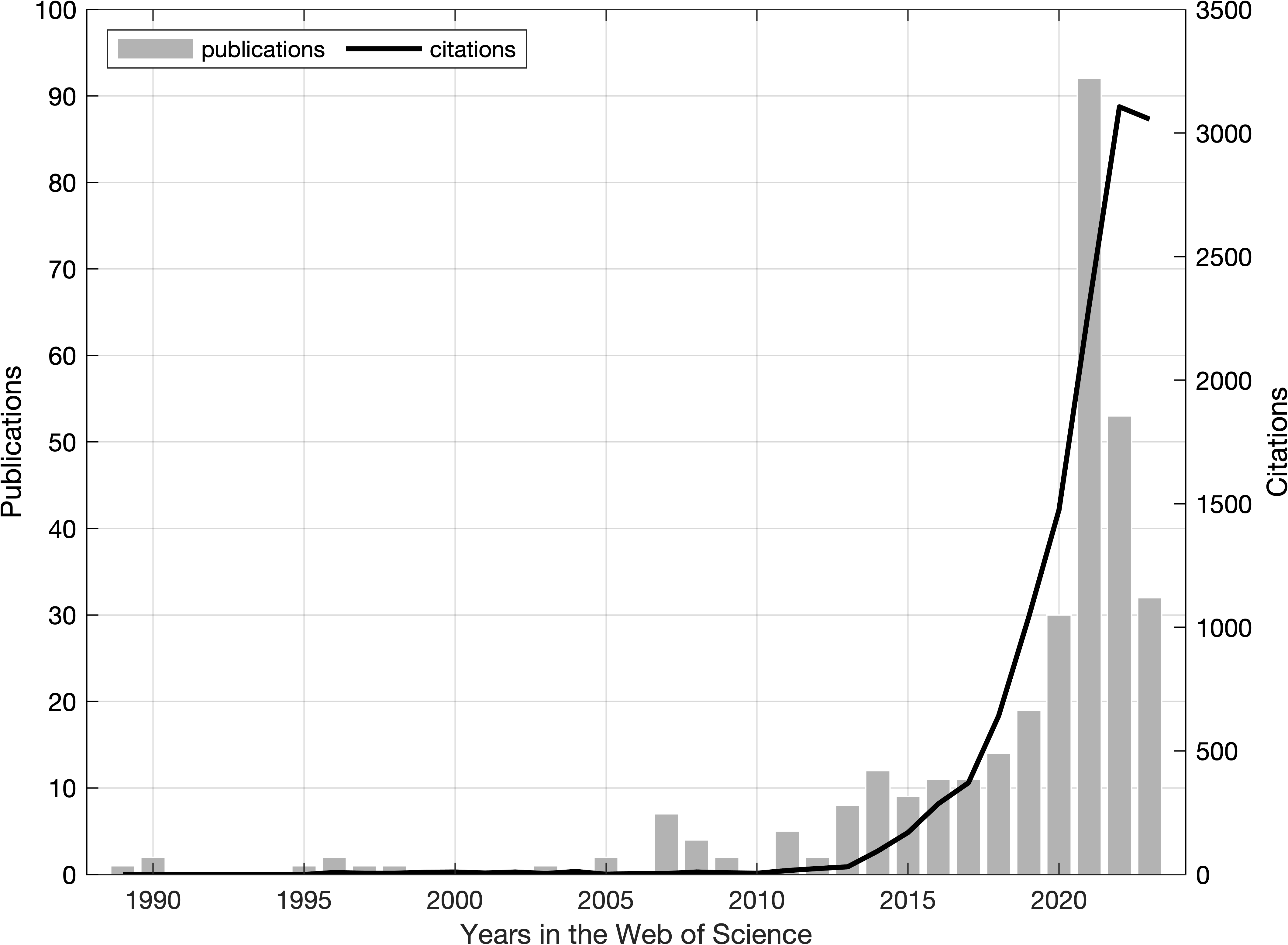}
\caption{Papermilling in plant science.}\label{fig:plant-science}
\end{figure}

\begin{figure}[t]
\center
\includegraphics[width=\columnwidth]{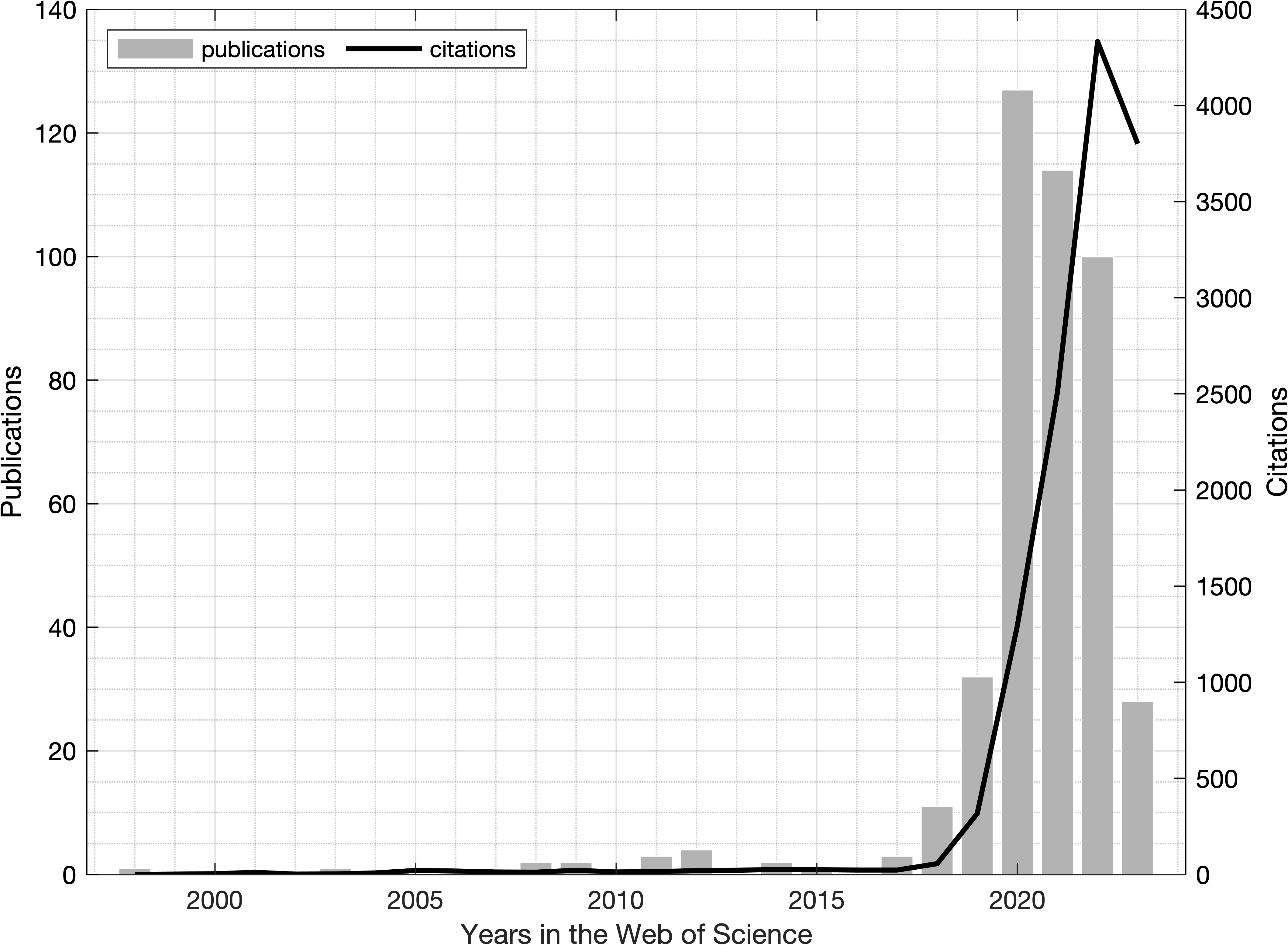}
\caption{Papermilling in engineering.}\label{fig:engineering}
\end{figure}

\section{Papermilling behavior \\ in other fields}

The exclusion of the field of mathematics from the Clarivate's lists of highly cited researchers
does not mean that other fields are spared  of papermilling, 
as the further two examples from Slovakia illustrate. 

In Fig.~\ref{fig:plant-science} a well-documented example of a researcher, listed by Clarivate in the past as highly cited in the field of plant science, is shown. 
This case brings the  picture similar to what we have seen in Fig.~\ref{fig:papermiller-example}, Fig.~\ref{fig:promotion1}, and Fig.~\ref{fig:promotion2}, namely  
 very low performance for long time in 1989--2013, and rapid growth after that: 
11 papers in 2017, 14 in 2018, 19 in 2019, 33 in 2020, peaking at 90 papers in 2021, accompanied by the growth of citation counts 
from roughly 30 citations  in 2013 to roughly 3100 citations in 2022.  
The number of co-authors is 1251, from 56 countries.

In Fig.~\ref{fig:engineering}, an example from the field of engineering is shown, 
very similar to the previous case, 
with very low performance until 2016, and extremely rapid growth after that:
3 papers and 23 citations in 2017,
11 papers and 56 citations in 2018,
32 papers and 307 citations in 2019,
more than 13 papers and around 1300 citations in 2020,
more than 110  papers and more than 2500 citations in 2021,
96 papers and more than 4300 citations in 2022. 
The number of co-authors is 1084, coming from 63 countries.  

We can conclude that the papermilling behavior in other fields is also widespread  
and has the same signs.

\section{\raggedright What made the papermilling behavior possible?}

It was the introduction of the APC (Article Processing Charges) and the payments for open access (OAF, open access fees), that radically changed the publication ecosystem 
and made the papermilling possible. 

Before the introduction of the APC, the publishers were getting their revenue by selling subscriptions to their journals and  reprints of the published papers. 
The authors provided their works to publishing houses for free or for a honorarium, which was a fraction of the revenue obtained by the publisher from selling the authors' works. 
In such a relationship, the publishers were interested in publishing the top-notch papers, 
because one  can hardly sell a meaningless or incorrect paper. 

With the introduction of APC, a new source of the publisher's revenue appeared:
a huge number of researchers wanting to publish their articles for the price of APC. 

The introduction of OAF made it possible for publishers to sign agreements with institutions and national authorities (such as ministries of education, science, research), in accordance with which those institutions or national authorities pay the OAF for the papers of their employees. 

In the APC/OAF publishing model, publishers are -- in the limit -- interested 
in making all authors  paying APC/OAF, because this will reduce to zero the risk of not selling any articles. 

The arguments that the APC and OAF support open access publishing are very questionable, because the Arxiv.org and similar repositories also provide open access. 
Even more, such repositories definitely provide the so called ``Diamond Open Access'', 
where the authors do not pay for publication and readers do not pay for reading. 

It should be mentioned that although the publishers receive the APC and OAF money from the authors or the authors' institutions, the reviewers of papers do their work for free -- they are not paid for their work. 
This creates an incentive for manipulating the reviewing process, and even to creation of the ``review mills'', 
as is documented in \cite{Oviedo-Garcia}.

\section{Conclusion}

Obviously, the evaluation of the quality of research performance 
cannot be based on a single number,
and it does not matter if 
that number is the h-index, 
or the number of publications, or the number of citations, 
or the integrity index proposed in this paper, or any other 
single quantitative indicator. 

However, there are some standard patterns typical for conscientious researchers, 
focused on the advancement of knowledge,
and for researchers exhibiting the papermilling behavior,
focusing on the numbers of publications, number of citations, and h-index,
which give them advantages and benefits.

The main conclusion is that the papermilling behavior is reflected 
in the strong correlation between publications counts and citations counts, 
without standard delay between publications and citations time series; 
in the persistent growth of the number of publications in one year,
leading to abnormally large number of articles in one year;
in the low integrity index, proposed in this article; 
in the increasing number of co-authors along with the increase of
the number of publications and citations; 
and with increasing number of countries where the co-authors reside. 
The simultaneous presence of several of these indicators 
means the high probability that a researcher is involved in 
some kind of papermilling, 
although there might be singular, or exceptional, cases, where the final 
conclusion after additional thorough examination can be different.

From the presented study it follows that, in addition to the recommendations 
listed in \cite{Agricola-etal-2025b}, the journal editors should pay attention 
not only to each submitted manuscript, avoiding the products of the paper mills, 
but also to the publication patterns of all its authors.

The detailed analysis of the dynamics of the changes of the number of co-authors 
related to the papermilling behavior requires the change of the format 
of the data exported from the Web of Science database 
(or eventually from the Scopus database, which also can be used for such studies).

\clearpage

\onecolumn

\section*{Addendum, Feb 2, 2026: 
Clarivate returned Mathematics  \\
	to the list  of Highly Cited Researchers, but...}

Clarivate returned the field of Mathematics to their HCR lists published on November 12, 2025, with an explanation provided in the FAQ section (\href{https://clarivate.com/highly-cited-researchers/faqs/}{https://clarivate.com/highly-cited-researchers/faqs/}).

\begin{quotation}
\noindent
\textit{
``This year we are reintroducing the Mathematics category after an absence of two years. There are 60 researchers named in Mathematics in 2025.'' 
}

%
\end{quotation}

\noindent
Let us take a close look at the data provided by Clarivate. Tidying up the list of 60 mathematicians can be done even manually, especially because of the medialization of the previous removal of the field of Mathematics from the HCR list. 
However, it looks like there is still a lot of headroom for improvement.

\vspace*{-2ex}

\section*{A1. Unverified identities}

Selecting ``Mathematics'' in the HCR list produces the list of 65 names -- \textit{not 60 names, as Clarivate informed} -- 
of which 48 names are listed with the ``View profile'' links to their personal profiles at the Web of Science. However,  17 names are listed with the ``Claim profile'' links, which means 
that the authors did not confirm that those profiles and those publications are theirs. 

This means that 26\% of the HCR list for Mathematics 
does not have a verified identity at the Web of Science. 
This is in contradiction with the introduction and use of authors' identifiers, such as 
Web of Science ResearcherID (former Publons), or \hbox{ORCID}. 
The affiliations can change, but the author's identity should remain the same. 
To make a correct list of publications and citations in the Web of Science across 
several primary and secondary affiliations over their careers, the authors have to merge the corresponding data in their personal profiles in the Web of Science, and this can be done only by them after logging into their accounts at Web of Science. 

It should not be a problem to contact, for example, 
Emmanuel Candes or Peter Scholtze, 
and ask them to create or verify their Web of Science profiles (which they still do not have as of January 28, 2026). 
Clearly, every researcher listed in the HCR list must have a verified profile.

\vspace*{-2ex}

\section*{A2. Highly cited researchers with zero publications \\ and zero citations?}\label{sec:zero-citations}

Additionally, even the ``View profile'' link sometimes produces questionable output.

\begin{figure}[t]
\center
\includegraphics[width=\columnwidth,height=0.75\columnwidth]{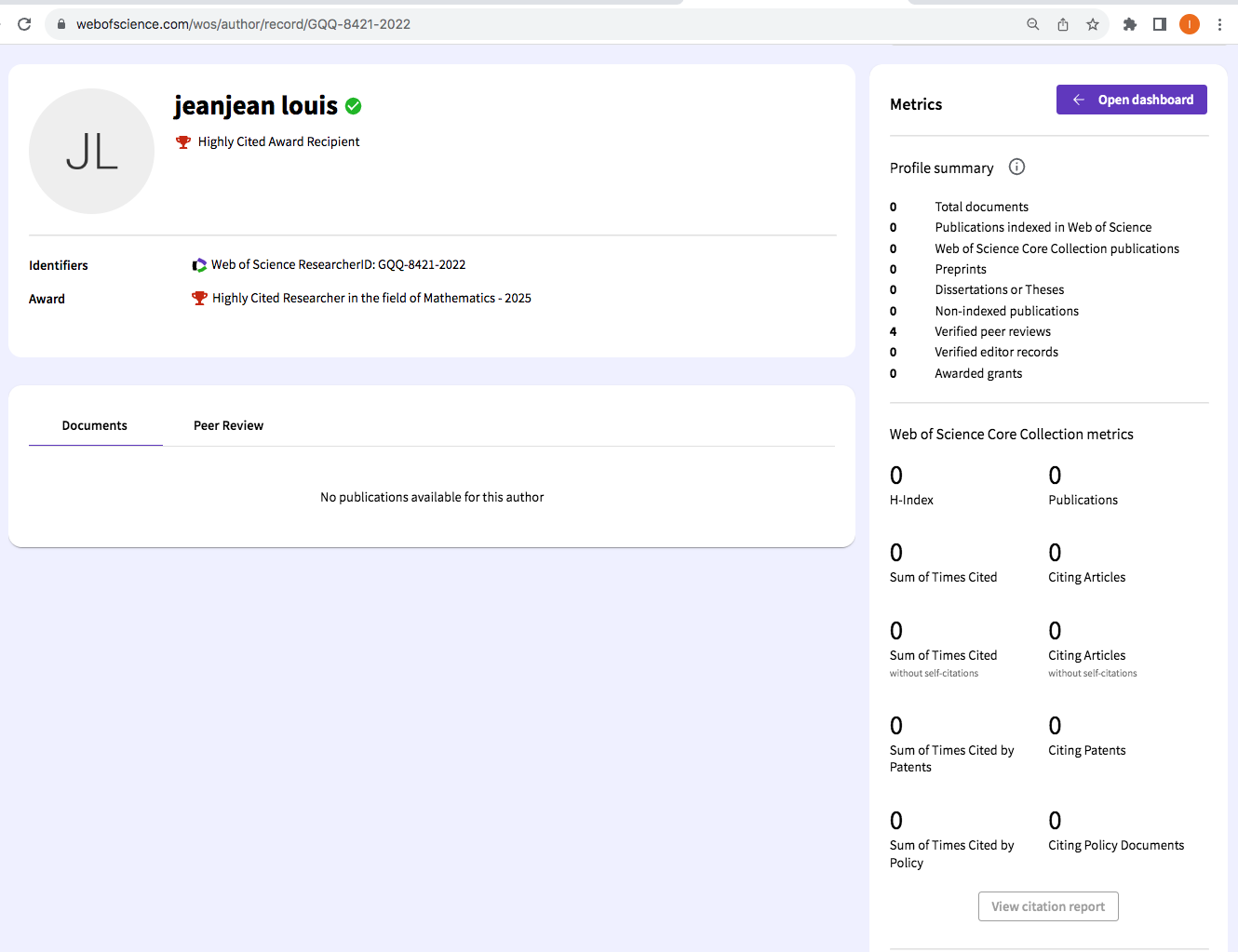}
\caption{Louis Jeanjean's profile in the Web of Science on Nov 17, 2025.}\label{fig:jeanjean1}
\end{figure}

\begin{figure}[t]
\center
\includegraphics[width=\columnwidth,height=0.75\columnwidth]{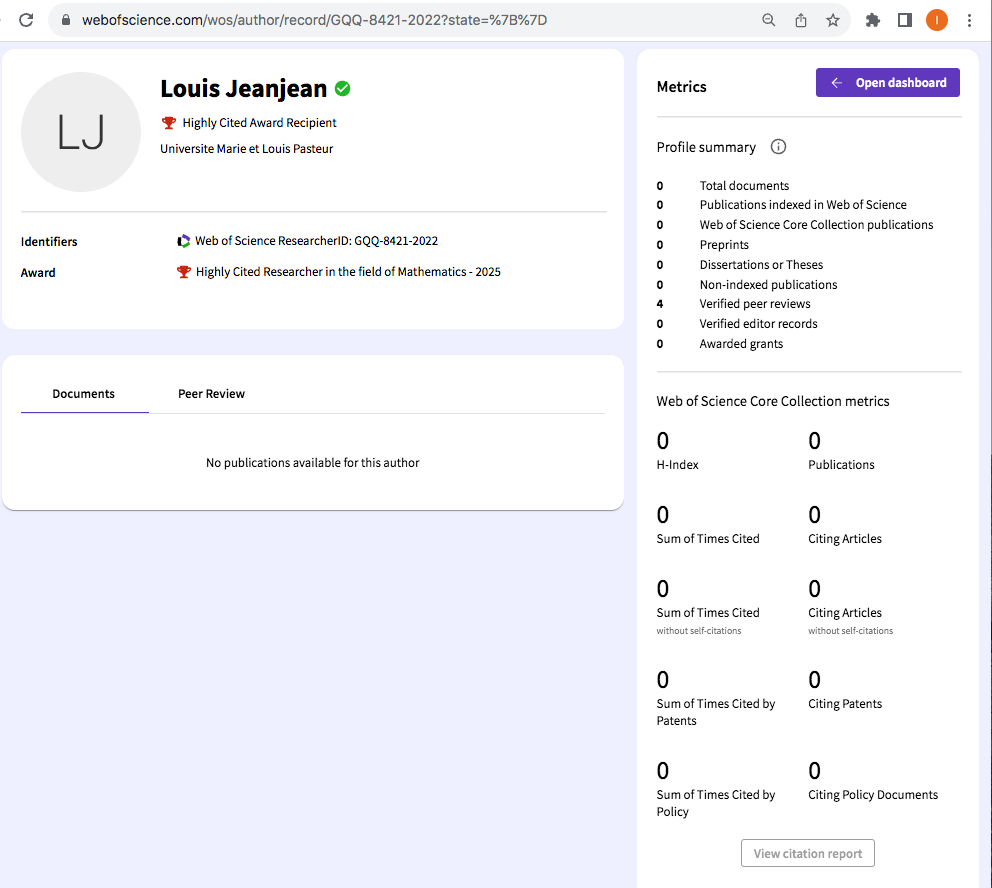}
\caption{Louis Jeanjean's profile in the Web of Science on Dec 08, 2025.}\label{fig:jeanjean2}
\end{figure}

The first example is Louis Jeanjean.
In the screenshot made on November 17, shown in Fig.~\ref{fig:jeanjean1}, 
the name is written using only lowercase letters, 
there is no affiliation provided, and all publication and citation numbers are zero -- and that all is strange. 
In the screenshot made on December 8, shown in  Fig.~\ref{fig:jeanjean2},
the name is capitalized, some affiliation is added, but  but the 
numbers of publications and citations are still all zero.

\begin{figure}
\center
\includegraphics[width=\columnwidth,height=0.75\columnwidth]{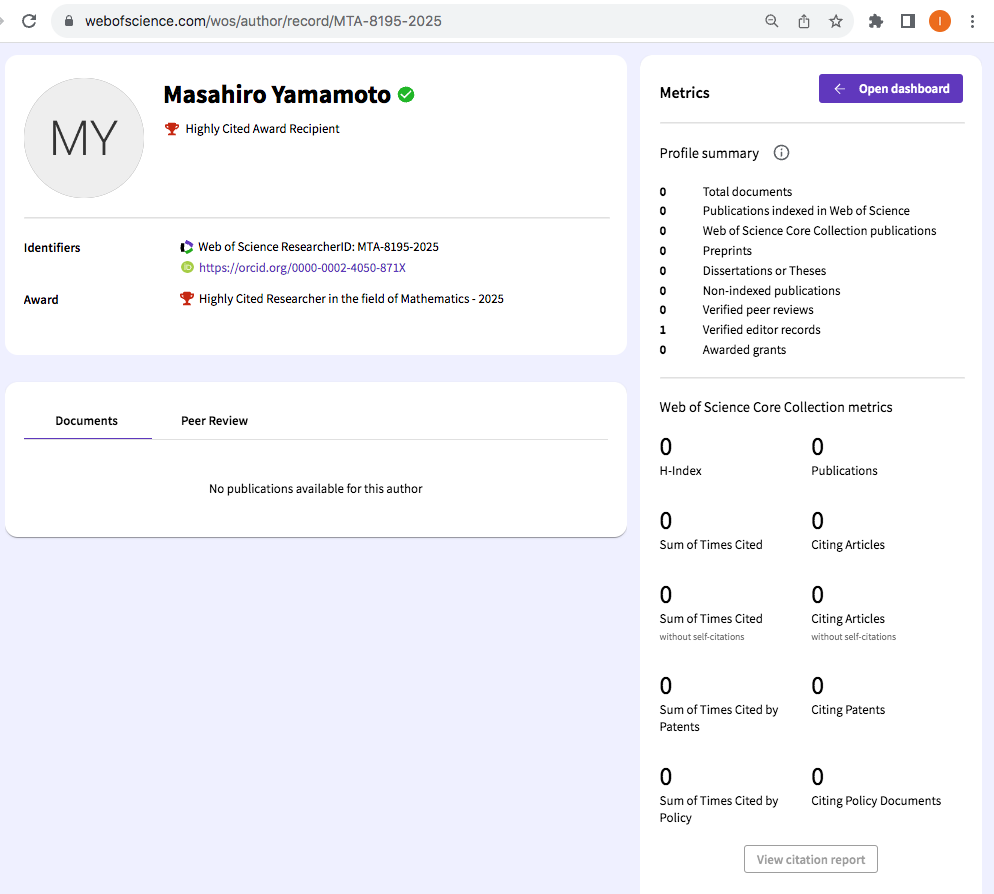}
\caption{Masahiro Yamamoto's profile in the Web of Science on Dec 8, 2025.}\label{fig:yamamoto2}
\end{figure}

The second example is Masahiro Yamamoto. 
In the online list, published by Clarivate, his affiliation is listed as ``Romanian Academy, Romania''.
However, there is not such institution, and the only similarly looking name is 
``Academy of Romanian Scientists'', which is  not an institution, but a learned society. 
The link ``View profile'' leads to the profile shown in Fig.~\ref{fig:yamamoto2}, 
without affiliation, and with zero papers. 

The search for Masahiro Yamamoto (and for Yamamoto Masahiro) in the Web of Science produces the list of 48 researchers. Two of them are HCRs, but they are not mathematicians. 
Three of those 48 are mathematicians, but they are not denoted as HCRs. 
It is not clear which one appears in the Clarivate's list of HCRs.

Both cases are unverifiable, and this is not the way of presentation of the persons
who Clarivate identified as highly cited mathematicians. 

\vspace*{-3ex}

\section*{\hbox{A3. No~East-European~names?}}\label{sec:no-East-Europeans}

This question arises as we notice that a mathematician Masahiro Yamamoto from the University of Tokyo is a member of previously mentioned ``Academy of Romanian Scientists''
(see \href{https://www.aosr.ro/membrii-sectiei-stiinte-matematice/}{https://www.aosr.ro/membrii-sectiei-stiinte-matematice/}). 
There is an unclaimed WoS profile with the ResearcherID OZT-6798-2025, 
with seven papers denoted as highly cited papers, 
where in the list of affiliations we see also ``Academy of Romanian Scientists'' 
Is this the correct Masahiro Yamamoto?

If yes, then it is unclear why a researcher with a Slavic name from Germany,
the regular co-author of Masahiro Yamamoto with the ResearcherID OZT-6798-2025,
is not included in the list of HCRs despite having twelve highly cited papers 
including three highly cited papers with (this) Masahiro Yamamoto.  

It is worth mentioning that, in general, the 2025 HCRs list does not contain 
researchers with names sounding Slavic or East European.

\section*{A4. Large collaborations in mathematics?}

Clarivate did not change the methodology for compiling the 2025 list of HCRs. 
They are still counting  the ``beans'' -- ``highly cited papers'', 
which are the papers that received some number of citations exceeding the thresholds. 
And the thresholds for Mathematics are really low:
to be marked as a highly cited paper,
3~citations are sufficient for paper published in 2025;
9~citations for a 2024 paper;
19, 30, 42, 56, 64, 74, 77, 81, 92~citations 
for papers published in 2023, 2022, 2021, 2020, 2019, 2018, 2017, 2016, 2015, correspondingly.

This led to including in the 2025 list of HCRs a~group of nine principal developers 
of  the \emph{deal.II} finite element software library, 
who all are co-authors on 
a series of eight papers -- one paper describing the design and features, 
and seven papers describing the improvements 
of the library (for versions 8.5, 9.0, 9.1, 9.2, 9.3, 9.4, 9.5), all with  similar abstracts: 
\textit{``This paper provides an overview of the new features of the finite element library deal.II, version [number]''.} 

Note that 9 names out of 65 means 14\%. 
Does this mean that mathematicians are pushed towards very questionable 
``large collaborations'' known in physics, medicine, pharmacology, and some other fields?
Just consider that the complete list of contributors to \emph{deal.II} 
currently contains 13 principal developers, 4~developers emeriti, 
and 418 contributors
(see \href{https://dealii.org/community/team/}{https://dealii.org/community/team/}).

\begin{figure}[t!]
\center
\includegraphics[width=\columnwidth,height=0.17\columnwidth]{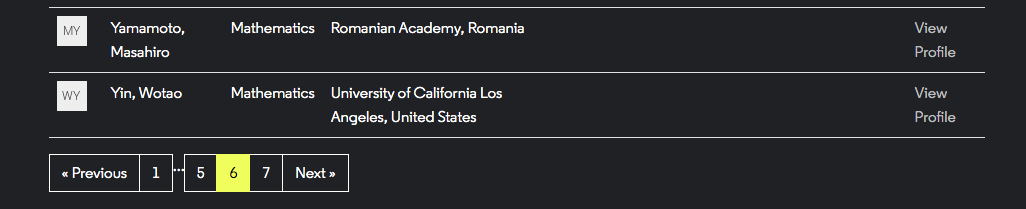}
\caption{Masahiro Yamamoto profile in the online HCR 2025 list on Dec 15, 2025.}\label{fig:yamamoto1}
\end{figure} 

\begin{figure}[t!]
\center
\includegraphics[width=\columnwidth,height=0.17\columnwidth]{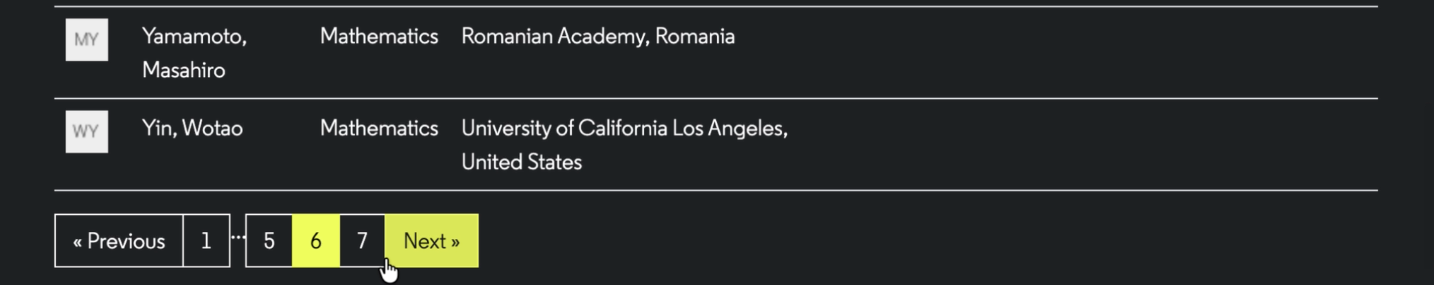}
\caption{Masahiro Yamamoto profile in the online HCR 2025 list on Dec 21, 2025. 
Notice that the links to ``View profile'' disappeared}\label{fig:yamamoto3}
\end{figure} 

\begin{figure}[t!]
\center
\includegraphics[width=\columnwidth,height=0.12\columnwidth]{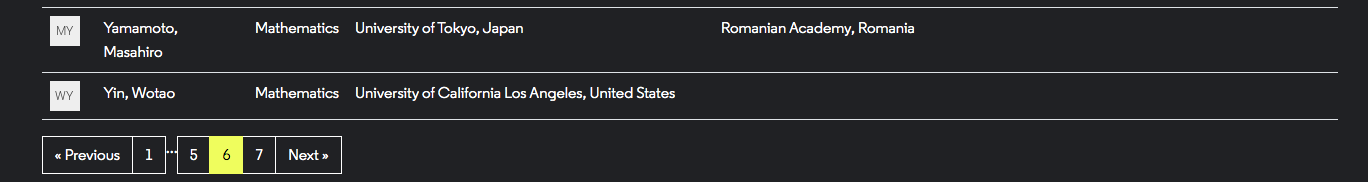}
\caption{Masahiro Yamamoto profile in the online HCR 2025 list on Jan 8, 2026. 
No ``View profile'' link. 
The University of Tokyo affiliation has been added, 
but the non-existent ``Romanian Academy'' is still there as the second affiliation.}\label{fig:yamamoto4}
\end{figure}

\section*{A5. Clarivate's silent updates of the HCR~2025 list}

I have written the above text of this Addendum on December 15,  2025. 
About a week later, on December 21, 
I have noticed that the HCR 2025 list has been silently updated, 
and the links to ``View profile'' and ``Claim profile'' have been removed -- 
compare the screenshots Fig.~\ref{fig:yamamoto1} and Fig.~\ref{fig:yamamoto3}. 
After this change, it became impossible to check out the profiles of 
the highly cited researchers in the Web of Science. 

Further, on January 8, 2026, I have noticed that 
the non-existent ``Romanian Academy'' is listed as Yamamoto's second affiliation,
and the primary affiliation ``University of Tokyo'' has been added. 
By this silent update, Clarivate in fact clarified the case of Masahiro Yamamoto 
in agreement with what I have written earlier in Section~\ref{sec:zero-citations} 
and in Section~\ref{sec:no-East-Europeans}. 

Even more: the  ``View profile'' link for Masahiro Yamamoto 
in the list published in the beginning of November, 2025, 
was leading to the profile 
of a material scientist with the same full name at the Konan University in Japan, 
as has been noted by Prof. Fatiha Alabau-Boussouira (see:
\href{https://retractionwatch.com/2025/11/12/math-is-back-as-clarivate-boosts-integrity-markers-in-highly-cited-researchers-list}{https://retractionwatch.com/2025/11/12/math-is-back-as-clarivate-boosts-integrity-markers-in-highly-cited-researchers-list}).

Then it remains unclear why  the aforementioned co-author of Masahiro Yamamoto, 
who has twelve highly cited papers compared with Yamamoto's seven highly cited papers, 
is not included in the HCR 2025, in contradiction with Clarivate's methodology.

\section*{Conclusion to the Addendum}

One can see that:

\begin{itemize}

\item
17 names (26\%) out of the total 65 did not have verified Web of Science profiles with claimed publications
at the time of publication of the HCR 2025 list;

\item
there were two names with zero publications and zero citations in their Web of Science profiles;

\item
a group of nine researchers (14\%) share eight papers on the regularly updated versions of a software package that they maintain;

\item 
the 2025 HCRs list does not contain 
researchers with names sounding Slavic or East European;

\item
the ongoing silent updates of the HCR 2025 list, including the removal of the ``View profile'' links, 
show that, indeed, this list was not curated properly. 

\end{itemize} 

The 2025 list of highly cited researchers in the field of mathematics 
is clearly very far from what it eventually could be. 
An attempt to fix the list of highly cited mathematicians 
by avoiding researchers exhibiting the papermilling behavior 
was not very successful.

\end{document}